\documentclass[twoside,a4paper,11pt]{sea10}
\usepackage{graphicx}
\usepackage[breaklinks=true]{hyperref}
\hypersetup{
    bookmarksopen=false,
    bookmarksnumbered=false,
    unicode=true,           
    pdftoolbar=true,        
    pdfmenubar=true,        
    pdffitwindow=false,     
    pdfstartview={FitH},    
    pdfnewwindow=true,      
    colorlinks=true,       
    linkcolor=red,          
    citecolor=magenta,        
    filecolor=green,      
    urlcolor=blue           
}

\usepackage{movie15}
\begin{document}
\pagenumbering{arabic}
\pagestyle{myheadings}
\thispagestyle{empty}
{\flushleft\includegraphics[width=\textwidth,bb=58 650 590 680]{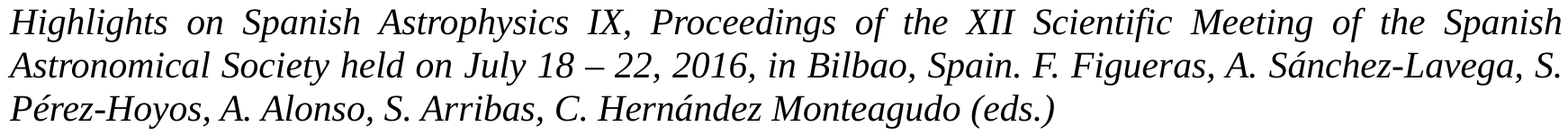}}
\vspace*{0.2cm}
\begin{flushleft}
{\bf {\LARGE
%
How much can we trust high-resolution spectroscopic stellar chemical abundances?
%
}\\
\vspace*{1cm}
%
S. Blanco-Cuaresma$^{1}$,
T. Nordlander$^{2}$,
U. Heiter$^{2}$,
P. Jofr\'e$^{3}$,
T. Masseron$^{3}$,
L. Casamiquela$^{4}$,
H. M. Tabernero$^{5}$,
S. S. Bhat$^{6}$,
A. R. Casey$^{3}$,
J. Mel\'endez$^{7}$,
and
I. Ram\'irez$^{8}$
%
}\\
\vspace*{0.5cm}
%
$^{1}$ Observatoire de Gen\`eve, Universit\'e de Gen\`eve, CH-1290 Versoix, Switzerland \\
$^{2}$ Dept. of Physics and Astronomy, Uppsala University, Box 516, 75120 Uppsala, Sweden \\
$^{3}$ Institute of Astronomy, University of Cambridge, Madingley Road, Cambridge, UK \\
$^{4}$ Dep. F\'isica Qu\`antica i Astrof\'isica, Universitat de Barcelona, ICCUB-IEEC, Spain \\
$^{5}$ Dpto. Astrof\'isica, Facultad de CC. F\'isicas, Universidad Complutense de Madrid, Spain \\
$^{6}$ Christ University, Hosur road, Bangalore- 560029, India \\
$^{7}$ Dept. de Astronomia do IAG/USP, Universidade de S\~ao Paulo; Brasil \\
$^{8}$ McDonald Observatory and Dept. of Astronomy, University of Texas at Austin; USA
%
\end{flushleft}
%
\markboth{
Trusting stellar chemical abundances
}{ 
%
S. Blanco-Cuaresma et al.
%
}
\thispagestyle{empty}
\vspace*{0.4cm}
\begin{minipage}[l]{0.09\textwidth}
\ 
\end{minipage}
\begin{minipage}[r]{0.9\textwidth}
\vspace{1cm}
\section*{Abstract}{\small
%
To study stellar populations, it is common to combine chemical abundances from different spectroscopic surveys/studies where different setups were used. These inhomogeneities can lead us to inaccurate scientific conclusions. In this work, we studied one aspect of the problem: When deriving chemical abundances from high-resolution stellar spectra, what differences originate from the use of different radiative transfer codes?
%
\normalsize}
\end{minipage}
%
%
%
\section{Introduction \label{intro}}

The imprints of the history of our Galaxy are kept in the stellar atmospheres \cite{2015A&A...577A..47B}, their chemical composition can help us to unravel past events that took place in the stellar aggregate where they belonged (or still belong) and in its surroundings. Abundances can also help to explain the nucleosynthesis since the abundance of certain elements can be altered by stellar evolution processes. This wide variety of topics has motivated an increase in the number of spectroscopic surveys in the last years, such as APOGEE \cite{2011AJ....142...72E} or the Gaia-ESO Public Spectroscopic Survey (GES) \cite{2012Msngr.147...25G} \cite{2013Msngr.154...47R}.

The increase in high-resolution spectra has lead to the development of tools for their automatic analysis \cite{2016arXiv160908092B}, each one with its peculiarities and its ingredients (e.g. model atmospheres, radiative transfer codes, normalization procedures). Therefore, the published abundances are not homogeneously derived and the scatter can be significant \cite{2014A&A...561A..93H}. On the other hand, it is common to compile chemical abundances from different studies to create bigger samples to increase the statistics \cite{2014AJ....148...54H}. But, given the mentioned inhomogeneities, this may affect the accuracy of the scientific conclusions based on combined results.

One of the ingredients in the spectroscopic pipelines that may lead to different results is the radiative transfer code. In \cite{2016arXiv160908092B} we presented an experiment to evaluate the impact on the determination of atmospheric parameters when different codes are used. In this study we evaluated the impact on the determination of chemical abundances using iSpec\footnote{\href{http://www.blancocuaresma.com/s/}{http://www.blancocuaresma.com/s/}} \cite{2014A&A...569A.111B}.

\section{Method}

We analyzed the Gaia FGK Benchmark Stars \cite{2014A&A...564A.133J} \cite{2015A&A...582A..81J} \cite{2015A&A...582A..49H} \cite{2016A&A...592A..70H}, a selection of very well-known stars accompanied with reference atmospheric parameters (i.e. effective temperature and surface gravity) derived independently from spectroscopy. We used the public high-resolution spectral library \cite{2014A&A...566A..98B} with a resolution of 47\,000, and we developed an automatic pipeline based on iSpec. The analysis was done using an atomic line list extracted from VALD \cite{2011BaltA..20..503K}, the MARCS\footnote{\href{http://marcs.astro.uu.se/}{http://marcs.astro.uu.se/}} model atmosphere \cite{2008A&A...486..951G}, solar abundances from \cite{2007SSRv..130..105G} and the following radiative transfer codes: SPECTRUM \cite{1994AJ....107..742G}, WIDTH9/SYNTHE \cite{1993KurCD..18.....K} \cite{2004MSAIS...5...93S}, SME \cite{1996A&AS..118..595V}, Turbospectrum \cite{1998A&A...330.1109A} \cite{2012ascl.soft05004P}, and MOOG \cite{2012ascl.soft02009S}.

We derive chemical abundances for 11 elements (i.e., Ca, Co, Cr, Fe, Mg, Mn, Ni, Sc, Si, Ti, V, although in this work we show only iron results) fixing the atmospheric parameters (i.e. effective temperature, surface gravity, metallicity, micro/macro-turbulence) to their reference value. The abundances were obtained not only using the synthetic spectral technique but also the equivalent width method since both are offered by iSpec.

\begin{figure}
\center
\includegraphics[scale=0.5]{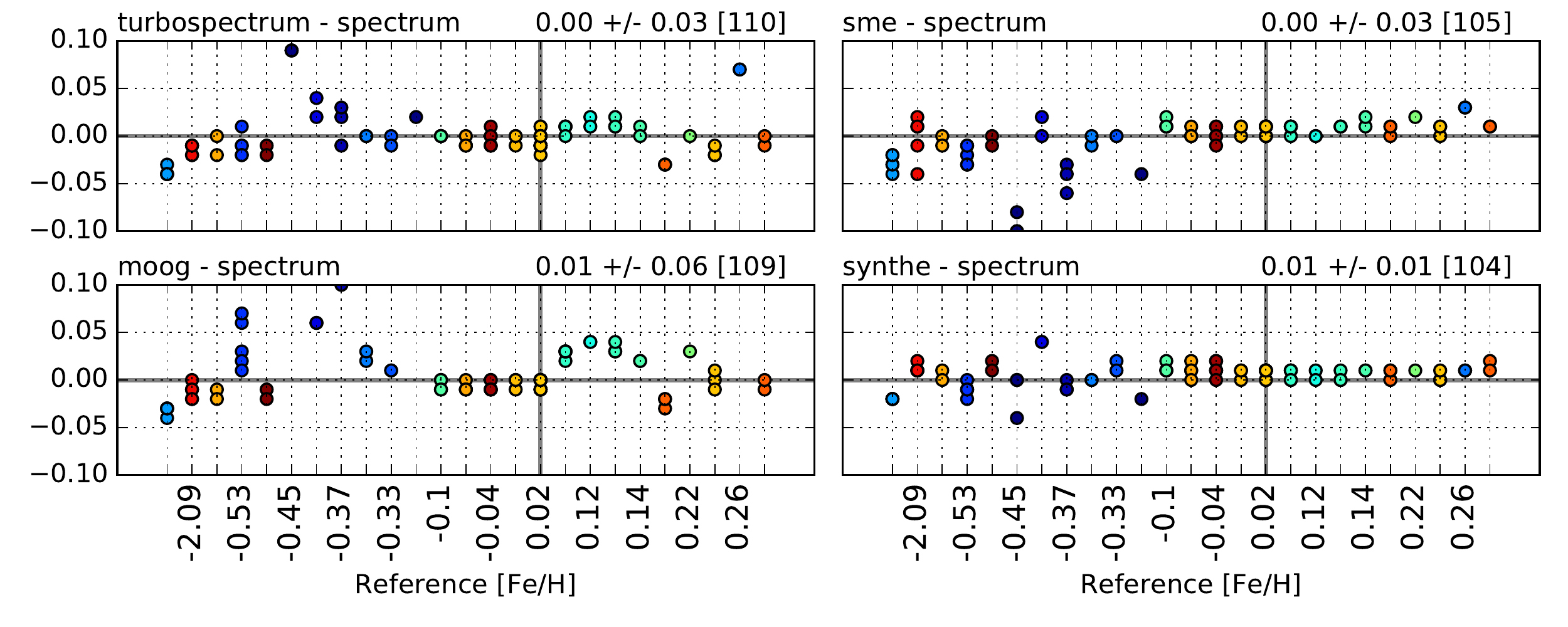}
\caption{\label{fig1} Differences in iron (synthesis) against SPECTRUM. Median difference and dispersion in the upper right text (mean number of lines in brackets). Effective temperature coded with colors (blue: cooler; red: hotter). Vertical gray line corresponds to the Sun.
}
\end{figure}

\section{Results}

Fig.~\ref{fig1} shows the differences in iron abundance (synthetic spectral technique) when comparing SPECTRUM to the rest of radiative transfer codes. The highest level of agreement (understood in terms of difference dispersion) is with SYNTHE, followed by SME and Turbospectrum. MOOG has the larger disagreement compared to SPECTRUM or any of the other codes.

\begin{figure}
\center
\includegraphics[scale=0.5]{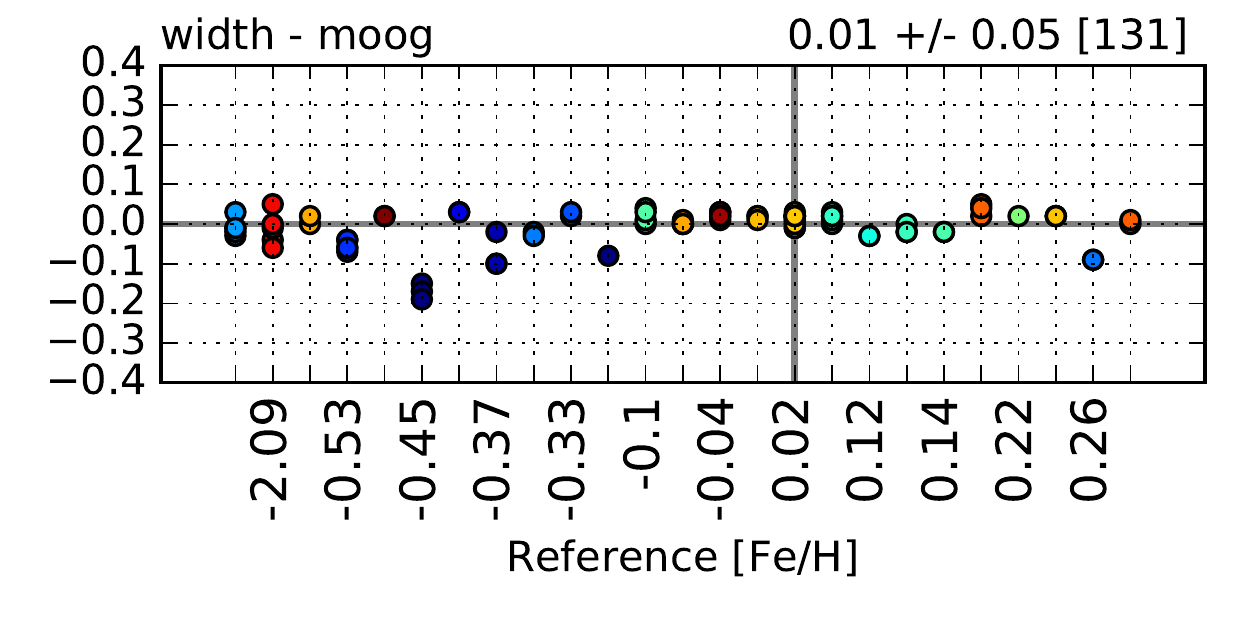}~
\includegraphics[scale=0.5]{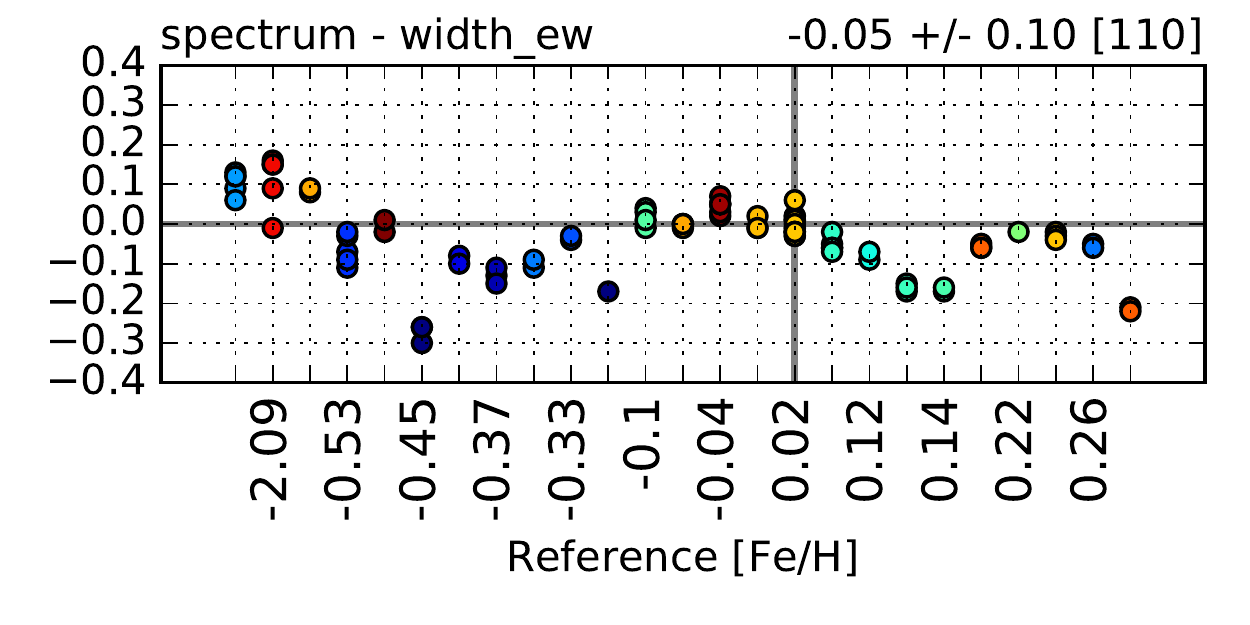}
\caption{\label{fig2} 
    Differences in iron abundance using only equivalent widths (left side), and synthesis compared to equivalent width (right side). Median difference and dispersion in the upper right text (mean number of lines in brackets). Effective temperature represented by colors (blue: cooler; red: hotter). Vertical gray line corresponds to the Sun.
}
\end{figure}

For the equivalent width method (left side in Fig.~\ref{fig2}), the level of agreement between the two radiative transfer codes is lower than the average agreement for synthesis. Note that the scale in the Y-axis was changed with respect to Fig.~\ref{fig1}. 

Finally, when synthesis and equivalent width abundances are compared (right side in Fig.~\ref{fig2}), the disagreement is even more relevant. It is important to remember that blends are taken into account when using the synthetic spectral technique, hence it is expected to have greater abundances when the equivalent width method is used.

\section{Conclusions}

Even if we have used the same atomic line list and model atmosphere for all the codes, not all of them use the same information. For instance, all the codes except MOOG recompute the electron number density internally, by solving the equation of state (i.e. the ionization fractions of different elements) consistently given the particular chemical composition. Additionally, MOOG is also the only radiative transfer code that does not accept any input parameter for the Stark broadening (it does an internal approximation). Other differences between synthesis codes may include continuous opacities and the treatment of scattering, the implementation of van der Waals broadening, sphericity effects, etc. Thus, every radiative transfer code has its peculiarities that leads to the differences presented in this study.

Even executing an homogeneous analysis with the same ingredients and the exactly same atmospheric parameters, we found differences in the determination of abundances that cannot be ignored. In \cite{2016arXiv160908092B}, we demonstrated how different codes also lead to differences in the determination of atmospheric parameters, we can expect that this will propagate to the determination of abundances and worsen the differences presented in this study. This results shows the importance of being extremely careful when combining chemical abundances derived by different surveys/studies with different pipelines and setups. If the option of re-analyzing all the spectra in an homogeneous way is not feasible, it is strongly recommended to assess the dispersion introduced and prove that its impact is not relevant for our scientific goal.

\small  
%
\section*{Acknowledgments}   
%
This work would not have been possible without the support of Laurent Eyer (University of Geneva). UH and TN acknowledge support from the Swedish National Space Board (Rymdstyrelsen).

%

%
\end{document}